# Water Stability and Nutrient Leaching of Different Levels of Maltose Formulated Fish Pellets

*[1]Keri Alhadi Ighwela, [2]Aziz Bin Ahmad and [1]A.B. Abol-Munafi*

[1]Faculty of Fisheries and Aqua-Industry, University Malaysia Terengganu (UMT), Malaysia
[2]Department of Biological Sciences, Faculty of Science and Technology (UMT), Malaysia

**Abstract:** The effects of different levels of maltose on feed pellet water stability and nutrient leaching were studied. Five treatments, including control with three replicates with setup (0.0, 20, 25, 30 and 35%). Pellet leaching rates were used to indicate pellet water stability. The results show that the presence of maltose in the diets significantly improved pellet water stability ($p<0.05$), but the leaching rates of the feed (35% maltose) observed higher than other feeds. Increased maltose resulted in the corresponding decrease in pellet stability. The protein leaching rate of control feed and feed (20% maltose) was significantly ($p < 0.05$) lower than the rates of other diets The lipid leaching rate of control feed was lower than the rates of other diets, while the feed (35% maltose) was more leaching rate. It improved feeds water stability is one important reason why maltose enhances fish growth.

## INTRODUCTION

The most important difference between aquatic animal feeds and land animal feeds is the durability of the aquatic feeds. Pellet water stability is an important quality parameter in the manufacture of aquaculture diets. High pellet water stability is defined as the retention of pellet physical integrity with minimal disintegration and nutrient leaching while in the water until consumed by the animal [1]. Many finfishes such as trout and catfish are instant feeders [2]. For these fish, the pellets need to be stable in the water for only a few minutes, while the shrimp needs a few hours [3]. The manipulation of feed particles by the animals during feeding further impedes the stability of the feed.

Published data concentrate on the characteristics and development of suitable binders [4, 1]. Little is known about the effect of feed components on water stability. Chen and Jenn [2] suggested that water stability, composition density and supplementation are the most important physical characteristics of aquatic animal feeds. Therefore, there is an urgent need for formulating suitable pelleted feeds, which incorporate cheap locally available materials. Although carbohydrates are non-essential dietary nutrients for fish, but they are the cheap source of dietary energy, which can be used to meet the entire metabolic energy requirements leaving proteins for growth [5-7]. However, there are instances of nutritional problems when excessive carbohydrates are fed to some species of fish [8]. In addition, they improve the pellet ability and water stability of feeds. The cereals are cheapest sources of dietary energy, in the form of carbohydrates in developing countries. Its contain 60% or more carbohydrate and oilseed meals 40% or more and these are often among the cheapest materials for tilapia feeds [9]. The objective of this study evaluated of maltose, which got from barley on water stability of test pellets and nutrient leaching in water such as protein and lipid.

## MATERIALS AND METHODS

**Feed Preparation:** Ingredient's formulations of the experimental diets are presented in Table 1. Barley maltose was added at 0, 20, 25, 30 and 35% in the diet. All feeds were formulated with ingredients commonly used, including fish meal, soya bean, wheat flour, cellulose, sunflower oil, mineral premix, vitamin premix, ascorbic acid, binder and chromium oxide. The mixture was paste well by machine with just enough water to obtain a soft

**Corresponding author:** Keri Alhadi Ighwela, Faculty of Fisheries and Aqua-Industry,
University Malaysia Terengganu (UMT), Malaysia.

Table 1: Proportions of different ingredients in the formulated feeds (% dry matter)

| Ingredient | Feed A (0.0% Mal) | Feed B (20% Mal) | Feed C (25% Mal) | Feed D (30% Mal) | Feed E (35% Mal) |
|---|---|---|---|---|---|
| Fish meal | 12 | 12 | 12 | 12 | 12 |
| Soya bean | 38 | 38 | 38 | 38 | 38 |
| Wheat flour | 10 | 10 | 10 | 10 | 10 |
| Maltose | 0 | 20 | 25 | 30 | 35 |
| Cellulose | 35 | 15 | 10 | 5 | 0 |
| Palm oil | 3 | 3 | 3 | 3 | 3 |
| Mineral premix | 0.5 | 0.5 | 0.5 | 0.5 | 0.5 |
| Vitamin premix | 0.5 | 0.5 | 0.5 | 0.5 | 0.5 |
| Vitamin C | 0.4 | 0.4 | 0.4 | 0.4 | 0.4 |
| Binder (CMC) | 0.5 | 0.5 | 0.5 | 0.5 | 0.5 |
| Chromic oxide | 0.1 | 0.1 | 0.1 | 0.1 | 0.1 |

Table 2: Mean ± S.D. Proximate composition and gross energy of the test feeds (% dry matter)

| Ingredient | Feed A (0.0% Mal) | Feed B (20% Mal) | Feed C (25% Mal) | Feed D (30% Mal) | Feed E (35% Mal) | L.S.D |
|---|---|---|---|---|---|---|
| Moisture | 8.86±0.43a | 8.39±1.44a | 9.22±0.52a | 9.82±1.12a | 9.62±0.05a | 1.37 |
| Protein | 33.27±0.87a | 33.70±0.44a | 33.85±0.50a | 33.56±1.83a | 33.27±0.88a | 1.94 |
| Lipid | 4.67±0.07a | 4.83±0.10a | 4.68±0.14a | 4.83±0.29a | 4.67±0.01a | 0.31 |
| Ash | 4.44±0.04a | 4.77±0.03a | 4.81±0.03a | 4.94±0.17a | 4.88±0.45a | 0.31 |
| Fiber | 13.62±1.19a | 11.23±0.17b | 8.93±0.36c | 8.71±0.35cd | 8.71±0.06d | 0.43 |
| NFE | 35.14±1.64c | 37.08±0.99ab | 37.91±0.87ab | 38.14±1.06b | 38.85±0.57a | 1.00 |
| Total energy (kJ g$^{-1}$) | 18.94±1.99bc | 18.66±0.23c | 19.17±0.34b | 19.67±0.85a | 19.26±0.46ab | 0.64 |

consistency. It was then extruded in the form of noodles, in a single layer, using a hand-operated noodle making machine, with 2.5 mm diameter perforations in the die. All the diets were oven dried at 60°C and were then broken manually to small sizes packed in heavy-duty plastic bags and stored at room temperature.

**Chemical Analysis of Diets:** The tested diets from each treatment were analyzed according to the standard methods [10] for dry matter, protein, lipid, ash and fiber. Dry matter was obtained after drying in an oven at 105°C for 24 h until constant weight. Nitrogen content was measured using Kjeldahl method using Kjeltec ™, 2100 FOSS and crude protein was estimated by multiplying nitrogen content by 6.25. Total lipids were determined by petroleum extraction. Ash was determined by incineration in a muffle furnace at 600°C for 6 hrs. Crude fibre was extracted by filter bag using an ANKOM Fiber Analyzer (Model No: ANKOM $^{200}$, Ankom Technology, Macedon, NY) described by Vansoest *et al.*, [11]. Nitrogen free extracted was calculated by difference the total percentages of moisture, crude protein, crude lipid, ash and crude fibre from 100% (Table 2). Total energy value was calculated from published values for the diet ingredients [5]. Samples were analyzed in triplicate.

**Determination of Stability of Pelleted Feeds:** The water stability of the pellets was determined over a period of 6 hours by wet durability test with considerable modification to suit the present situation. Triplicate 5 grams samples of pellet of each diet were dropped into 15 glass beakers, which contained 800 ml tap water. The immersion times examined were 1h, 2h, 4h and 6h respectively. After immersion, the un dissolved solids and water were filtered through filter paper and were dried in the oven (105°C for 30 min), followed by further drying at 65°C to a constant weight, then cooled in a desiccators. The mean differences in weights of beakers containing the feed before immersion and after drying were used to calculate the percentage dry matter loss, which is a measure of the water stability of the pellets for the corresponding time intervals.

$$\text{Leaching rate} = \frac{A \times (1-r) - R}{A \times (1-r)} \times 100$$

where,

A = Weight of pellets before immersion;
r = Moisture content of pellets; and
R = Dry weight of the remaining solid.

**Protein and Lipid Leaching in Diets:** After the determination of pellet water stability, leaching of protein and lipid were determined by the microkjeldahl and soxhlet extraction of samples' methods of AOAC (1990) and expressed on a percentage remaining basis as follows:

Leaching of total protein

$$\text{or lipid (\% remaining)} = \frac{\text{g protein or lipid remaining/ g pellet remaining}}{\text{g protein or lipid nutrient/ g initial pellet}} \times 100$$

**Statistical Analysis:** All data obtained on the chemical analysis of diets. Water stability and nutrient leaching were subjected to the analysis of variance (ANOVA) and the differences in means were tested for significant using the Duncan multiple range test at 95% confidence level [12]. The analysis was supported by correlation two tailed using Genstat5 program.

## RESULTS

The proportions of ingredients used in the formulation of feeds were based on the proximal analysis of the dry ingredients with a view to maintaining the protein, lipid and carbohydrate contents in all types of pellets at different levels of maltose each, which is considered suitable for rapid growth of warm water fish [13].

**Water Stability:** The results of the water stability test of experimental diets are presented in Table 3. At the initial of 15 min soaked in water, all diets recoveries were stable more than 82%. Water stability was increased with the decreases the percentage of maltose in diets. On the other hand, water stability diet was decreased by the increase the immersion period (4 h). While at the end of 6 hrs, the water stability of all diets was found to be almost the same. Feed 'E' was slighted stable compared to other diets (68%) at this condition.

**Nutrient Leaching Tests:** The nutrient retentions tests of experimental diets are given in Table 4 and Table 5. The nutrient retentions at the initial 15 min were, 97.57 to 99.01% for proteins and 96.27 to 98.92% for lipids, while at the end of 6 hrs the crud lipid retention of feed 'A' had highest (76.28%) compared to other diets. The feed 'B' and 'C' was no significant difference at all times of immersion. The feed 'E' it had less retention than other diets (70.75%). The crude protein retention after 6 hrs was not significantly different among diets. While at the end of 4hrs; the feed 'D' and 'E' had higher retention compared to other diets, but no significant difference at the 1h and 2hrs of immersion.

Table 3: Mean percentage water stability of pelleted feeds bound with maltose in freshwater

| Ingredient | Feed A (0.0% Mal) | Feed B (20% Mal) | Feed C (25% Mal) | Feed D (30% Mal) | Feed E (35% Mal) | L.S.D |
|---|---|---|---|---|---|---|
| 15 minute | 94.66±1.15a | 90.00±2.00b | 88.00±2.00b | 85.00±1.00c | 82.00±0.00d | 2.70 |
| 30 minuets | 93.33±1.15a | 87.33±1.15b | 84.66±0.57c | 83.00±0.00d | 80.00±0.00e | 1.49 |
| 1 hour | 91.33±0.15a | 84.60±0.40b | 81.13±0.30c | 79.13±0.82d | 77.83±0.25e | 0.81 |
| 2 hours | 85.93±0.41a | 82.96±0.51b | 79.43±0.15c | 77.50±0.51d | 74.30±0.10e | 0.64 |
| 4 hours | 80.23±0.40a | 78.33±0.15b | 76.36±0.25c | 74.83±0.25d | 72.03±0.30e | 0.55 |
| 6 hours | 77.50±0.26a | 75.86±0.20ab | 74.00±0.26ab | 71.80±0.26bc | 68.43±5.86c | 4.97 |

Table 4: Mean percentage of lipid remaining of pelleted feeds in freshwater

| Ingredient | Feed A (0.0% Mal) | Feed B (20% Mal) | Feed C (25% Mal) | Feed D (30% Mal) | Feed E (35% Mal) | L.S.D |
|---|---|---|---|---|---|---|
| 15 minute | 99.01±0.19a | 98.05±0.13b | 97.82±0.19bc | 97.63±0.06c | 97.56±0.19c | 0.346 |
| 30 minuets | 97.50±0.13a | 96.10±0.13b | 96.46±1.16ab | 94.89±0.19c | 92.76±0.65d | 1.086 |
| 1 hour | 96.93±2.01 | 93.94±2.70 | 95.69±1.29 | 86.96±7.53 | 87.28±4.23 | 8.48 |
| 2 hours | 94.30±2.01 | 86.15±12.74 | 82.76±1.29 | 81.73±3.28 | 82.45±6.49 | 11.32 |
| 4 hours | 64.47±1.31c | 67.09±2.70bc | 64.65±3.41c | 73.91±3.98a | 72.81±4.02ab | 6.480 |
| 6 hours | 54.82±13.69 | 60.60±7.49 | 48.27±13.94 | 57.39±1.30 | 47.36±6.96 | 20.20 |

Table 5: Mean percentage of protein remaining of pelleted feeds in freshwater

| Ingredient | Feed A (0.0% Mal) | Feed B (20% Mal) | Feed C (25% Mal) | Feed D (30% Mal) | Feed E (35% Mal) | L.S.D |
|---|---|---|---|---|---|---|
| 15 minute | 98.92±0.57a | 97.70±0.49ab | 97.07±0.09bc | 96.46±1.36bc | 96.27±0.10c | 1.369 |
| 30 minuets | 97.76±0.41a | 95.58±1.52b | 95.40±0.41b | 94.57±0.04b | 94.29±0.56b | 1.474 |
| 1 hour | 95.77±0.01a | 89.21±0.19b | 89.00±0.32b | 88.53±0.74b | 86.98±0.43c | 0.814 |
| 2 hours | 89.42±0.10a | 85.17±0.48b | 84.89±0.76b | 83.24±0.18c | 82.23±0.57d | 0.643 |
| 4 hours | 84.61±0.78a | 79.88±0.82b | 79.64±1.31bc | 77.97±0.69cd | 76.57±0.99d | 1.878 |
| 6 hours | 76.28±0.59a | 75.10±0.46b | 74.74±0.54b | 72.69±0.36c | 70.75±0.46d | 1.008 |

## DISCUSSION

In feed formulation, water stability and nutrient leaching are the main issues. Finding showed that feed added with maltose had high-water stability. Although the leaching rate, but it is lower compared to time taken by fishes to consume the feed; 10 to 15 minutes [14]. Therefore the nutrient leaching due to maltose is acceptable and not considered as a major factor. The percentage of maltose added to the significantly contributed to the stability of feed (Table 3). For the first six hours, feed E was the less stable (68.43±5.86) than control. The control feed has good water stability might be due to the presence of cellulose and wheat flour as the binder [15, 16]. Feed E contained only maltose. Highly polarity or interaction energy between the polysaccharide chain and the ionic liquids improves water stability of feed pellets [17]. The degree of feed stability is directly related to the degree of gelatinization during steam conditioning [18]. The maltose added feed also demonstrated low leaching percentage of lipid and protein (Table 4 and Table 5). Feed 'E' has 20% less than control which is accordance with previous findings by Jayaram and Shetty [18] and Meyer [19]. Meyer [19] indicated that feed pellets can lose more than 25% of their crude protein and lipid content during two-hour immersion in water. The successful decrease in crude lipid and crude protein retention at increasing times of immersion represents the loss of free fatty acids and free amino acids present in the diets as a result of disintegration and leaching. Moisture facilitated the breakdown of proteins and loss of amino acids inform of Nitrogen (N) [20, 21]. Protein molecules had a large proportion of the subunits are mutually linked by disulfide bonds [22]. In addition, the addition of fat to the surface of pellets may also improve palatability of the diet and water stability and reduce the leaching rate of water soluble nutrients [23].

## CONCLUSION

The result confirmed that there are great potentials in maltose, which got from barley in fish feed stability and nutrients retention. The results are the preliminary findings in series of Ph. D work considering some biochemical characteristics of feedstuffs and additives in the development of improved diets for aquaculture.